\documentclass{article}

\PassOptionsToPackage{numbers, compress}{natbib}

\usepackage{ai3d_arxiv}

\usepackage[utf8]{inputenc} 
\usepackage[T1]{fontenc}    
\usepackage{hyperref}       
\usepackage{url}            
\usepackage{booktabs}       
\usepackage{amsfonts}       
\usepackage{nicefrac}       
\usepackage{microtype}      
\usepackage{xcolor}         

\usepackage{amsmath}
\usepackage{multirow}
\usepackage{graphicx}

\usepackage{makecell}

\title{Coordinate-Queryable Neural Field Reconstruction for EEG Spatial Super-Resolution with Unseen-Electrode Generation}
\author{
\begin{tabular}{c}
Hongjun Liu\textsuperscript{1,*} \quad
Leyu Zhou\textsuperscript{1,*} \quad
Zijianghao Yang\textsuperscript{1} \quad
Chao Yao\textsuperscript{2,\ensuremath{\dagger}} \\[0.35em]
\textsuperscript{1}School of Intelligence Science and Technology, University of Science and Technology Beijing \\
\textsuperscript{2}School of Computer and Communication Engineering, University of Science and Technology Beijing
\end{tabular}
}

\begin{document}

\maketitle
\begingroup
\renewcommand{\thefootnote}{\fnsymbol{footnote}}
\footnotetext[1]{Equal contribution.}
\footnotetext[2]{Corresponding author.}
\endgroup

\begin{abstract}
EEG spatial super-resolution (EEGSR) in real deployments is challenged by random channel missingness, unstable electrode quality, and changing visible-channel patterns caused by bad contacts or device variability. Most existing EEGSR methods learn a fixed low-to-high channel mapping under pre-defined input-output layouts, which makes them brittle when missing channels vary at test time. In this paper, we reformulate EEGSR as learning a shared conditional scalp field from partially observed support channels. Specifically, a position-guided encoder summarizes the observed EEG channels and their coordinates into a latent condition, and a conditional implicit neural representation decoder reconstructs target EEG signals by querying this condition at desired electrode coordinates. During inference, the model directly reconstructs unseen electrode signals from the available EEG support and the queried coordinates. To strengthen the constraint of the encoded latent representation on the decoder and thereby construct a more stable scalp field consistent with the observed channels, we further introduce a fidelity-preserving channel corruption training strategy under mixed electrode states. Extensive experiments across multiple EEG datasets demonstrate the effectiveness of our framework for both random missing-channel reconstruction and strict unseen-electrode signal generation. Notably, under the strict held-out-electrode setting on AAD, our method reduces NMSE by 37.5\% and improves SNR by 2.12 dB over the strongest baseline, showing its ability to synthesize signals at electrode locations never exposed during training.
\end{abstract}

\section{Introduction}

Electroencephalography (EEG) is widely used in brain--computer interfaces \citep{lawhern2018eegnet}, neural monitoring \citep{nejedly2019deep}, and clinical assessment \citep{jaramillo2023spectral} because it provides a practical and non-invasive way to capture brain activity. In many real deployment scenarios, however, high-density (HD) EEG acquisition remains difficult due to hardware cost, preparation burden, and limited wearing convenience \citep{wang2025generative}. As a result, practical systems often rely on low-density (LD) EEG signals, which substantially reduces spatial fidelity and may distort downstream decoding and visualization. EEG spatial super-resolution (EEGSR) aims to recover HD signals from sparse observations, offering a promising way to bridge practical acquisition and high-quality spatial analysis.

Existing EEGSR methods can be broadly grouped into three lines. Early studies mainly relied on interpolation-based reconstruction, which estimates missing channels from spatial smoothness priors or neighboring electrode relationships \citep{luo2020eeg}. More recent learning-based approaches replace handcrafted interpolation with data-driven low-to-high mapping, using convolutional, graph-based, or Transformer architectures to directly regress HD EEG from LD observations \citep{8333379,kwon2019super,li2025estformer}. More recently, generative models, including GAN- and diffusion-based methods, have further been introduced to improve reconstruction realism and spectral fidelity by modeling the distribution of HD EEG conditioned on sparse inputs \citep{wang2025generative,liu2026step}. These advances have substantially improved reconstruction quality under pre-defined experimental settings and demonstrated the potential of learning-based EEGSR for practical sparse-sensing scenarios.

Despite recent progress, most existing EEGSR methods are still built around pre-defined low-to-high channel mappings under fixed input-output electrode layouts \citep{8333379,kwon2019super, wang2025generative,liu2026step}. Such a formulation becomes fragile in realistic deployment, where electrode quality may vary across sessions or devices due to imperfect contact, unstable impedance, motion-related artifacts, or bad channels \citep{leach2020protocol,moumane2024signal}. When the missing channels change at test time, these methods can easily fall back to channel-index-specific correlations rather than learn a reconstruction rule that is truly grounded in scalp geometry. Classical interpolation methods provide a geometry-aware alternative \citep{luo2020eeg}, but they tend to oversmooth fine-grained spatial structure and offer limited flexibility when the target signals need to be reconstructed at electrode locations that are not explicitly observed during training. These limitations suggest that EEGSR under random channel missingness should not be treated as restoring a fixed output tensor from a fixed masked input tensor. Instead, it is more naturally viewed as reconstructing a coordinate-queryable scalp field. 

Following this perspective, we propose ScalpINR, a support-conditioned implicit neural representation framework for EEGSR under random channel missingness. Instead of learning a fixed low-to-high channel mapping, ScalpINR learns a conditional scalp field from partially observed support channels and reconstructs EEG signals by querying arbitrary target electrode coordinates. Specifically, a position-guided channel encoder summarizes the available channel signals and their spatial coordinates into a latent condition, while a coordinate-queryable implicit neural representation decoder predicts target signals at queried locations. To prevent the queried reconstructions from drifting away from the observed evidence, we further introduce a fidelity-preserving channel corruption strategy, which trains the model under mixed visible, corrupted, and missing electrode states and strengthens consistency with observed support channels. 
Our contributions are summarized as follows:
\begin{itemize}
    \item We reformulate EEG spatial super-resolution under random channel missingness as a support-conditioned coordinate-queryable scalp reconstruction problem, which unifies variable missing-channel reconstruction and strict unseen-electrode signal generation within the same framework.

    \item We propose ScalpINR, a position-guided implicit neural representation framework that encodes partially observed EEG support channels and their electrode coordinates into a latent condition, and reconstructs target signals by querying a conditional INR decoder at arbitrary electrode locations.

    \item We introduce a fidelity-preserving channel corruption training strategy that trains the model under mixed visible, corrupted, and missing electrode states, encouraging the inferred scalp field to remain faithful to observed support channels and improving reconstruction stability under sparse and irregular observations.
\end{itemize}

\section{Related Works}

\subsection{EEG spatial super-resolution and channel reconstruction}
Recent learning-based EEGSR methods typically train an end-to-end mapping from a fixed low-density (LD) montage to a fixed high-density (HD) montage under a pre-defined upsampling factor. Representative architectures include CNN or Transformer feature upsampling \citep{li2025estformer}, as well as generative paradigms such as GAN-based SR and diffusion-based conditional generation \citep{wang2025generative}. More recent diffusion variants incorporate richer spatial inductive biases, such as topology-aware embeddings and learned channel-relation graphs \citep{yao2026geometry}, or step-aware conditioning strategies \citep{liu2026step}. Despite improved fidelity, these approaches predominantly operate on fixed input/output electrode sets and remain coupled to a pre-defined channel inventory, which limits robustness when the visible channel set changes at test time.

A small number of recent works relax the fixed-montage assumption. ZUNA introduces a large-scale masked diffusion autoencoder with 4D positional encoding, enabling inference on arbitrary channel subsets and electrode positions \citep{warner2026zuna}.  
In contrast to these lines, our work enforces an explicit manifold spectral field parameterization that makes coordinate querying analytically well-defined and enables principled control of non-identifiable high-frequency components under severe missingness.

\subsection{Implicit neural representations for time-series and EEG neural fields}

Implicit neural representations (INRs) have recently been adapted to time-series to model signals as continuous functions of time, supporting irregular sampling and fine-grained imputation. TimeFlow proposes a continuous-time framework for imputation and forecasting based on conditional INRs and meta-learning-driven modulation \cite{naour:hal-04759780}. ImputeINR applies INR-based continuous modeling to clinical time-series imputation and reports gains especially under high missing ratios \citep{li2025imputeinr}. Beyond deterministic INRs, probabilistic variants such as temporal variational INRs learn distributions over continuous-time generator functions for individualized imputation and forecasting \citep{koyuncu2025temporal}. These methods mainly focus on temporal irregularity and point-wise missingness, and typically treat multivariate dimensions as fixed channels without explicitly modeling sensor geometry.
More closely related to our setting, Neural Brain Fields \citep{kedem2025neural} models EEG as a coordinate-queryable continuous neural field and supports rendering signals at previously unseen electrode positions. However, NBF fits a separate neural field for each recording, aiming at sample-specific continuous representation learning and novel electrode rendering, rather than learning a shared conditional reconstruction model under random channel missingness.

In contrast, our method targets EEG spatial reconstruction from partially observed support channels. We learn a shared conditional INR that infers a support-consistent scalp field from randomly missing EEG observations and reconstructs target signals at query coordinates, thereby jointly addressing variable missing-channel inputs and unseen-electrode generation under a unified reconstruction framework.



\section{Method}

Given a high-density EEG segment $X \in \mathbb{R}^{C \times T}$ with $C$ electrodes and $T$ time samples, we consider the realistic setting where only a subset of channels is available due to random channel missingness or unstable electrode quality. Let $\mathcal{L}$ denote the visible channels and $\mathcal{U}$ the unavailable ones, with $\mathcal{L}\cup\mathcal{U}=\{1,\dots,C\}$ and $\mathcal{L}\cap\mathcal{U}=\emptyset$. Our goal is not only to reconstruct signals on missing channels, but also to support signal generation at unseen electrode locations that are never provided as input during training. To this end, we formulate EEG spatial super-resolution as a coordinate-queryable scalp reconstruction problem from randomly missing channel observations. As illustrated in Fig.~X, our framework contains two core components: a \emph{position-guided channel encoder}, which summarizes the spatial layout and signal context of the available channels, and a \emph{coordinate-queryable conditional INR decoder}, which reconstructs EEG signals by querying the inferred scalp field at arbitrary electrode coordinates. To further prevent weakly constrained generation, we introduce a \emph{fidelity-preserving channel corruption training} strategy that enforces consistency between reconstructed signals and the already observed channels.

\subsection{Position-guided channel encoder}

A central challenge in EEGSR under random missingness is that the visible channels may vary across samples, sessions, or devices. Instead of assuming a fixed low-density input layout, we explicitly encode the spatial configuration of the available channels. For each visible channel $c \in \mathcal{L}$, we associate its EEG signal $x_c \in \mathbb{R}^{T}$ with its electrode coordinate $p_c \in \mathbb{R}^{3}$. The coordinate is transformed by a position embedding function $\phi(\cdot)$ to obtain a spatial descriptor:
\begin{equation}
e_c = \phi(p_c).
\end{equation}
We then combine the signal representation and position embedding to form a channel-wise token:
\begin{equation}
h_c = \Psi(x_c, e_c),
\end{equation}
where $\Psi(\cdot)$ denotes a lightweight feature projection module. The collection of visible-channel tokens is fed into a channel encoder $E_{\theta}$ to aggregate spatial cues from randomly missing EEG observations:
\begin{equation}
z = E_{\theta}(\{h_c \mid c \in \mathcal{L}\}),
\end{equation}
where $z$ denotes the latent scalp representation conditioned on the currently available channels. In this way, the encoder learns to summarize both signal context and electrode layout from variable visible-channel patterns, rather than from a fixed input slot arrangement.

\begin{figure}
    \centering
    \includegraphics[width=\linewidth]{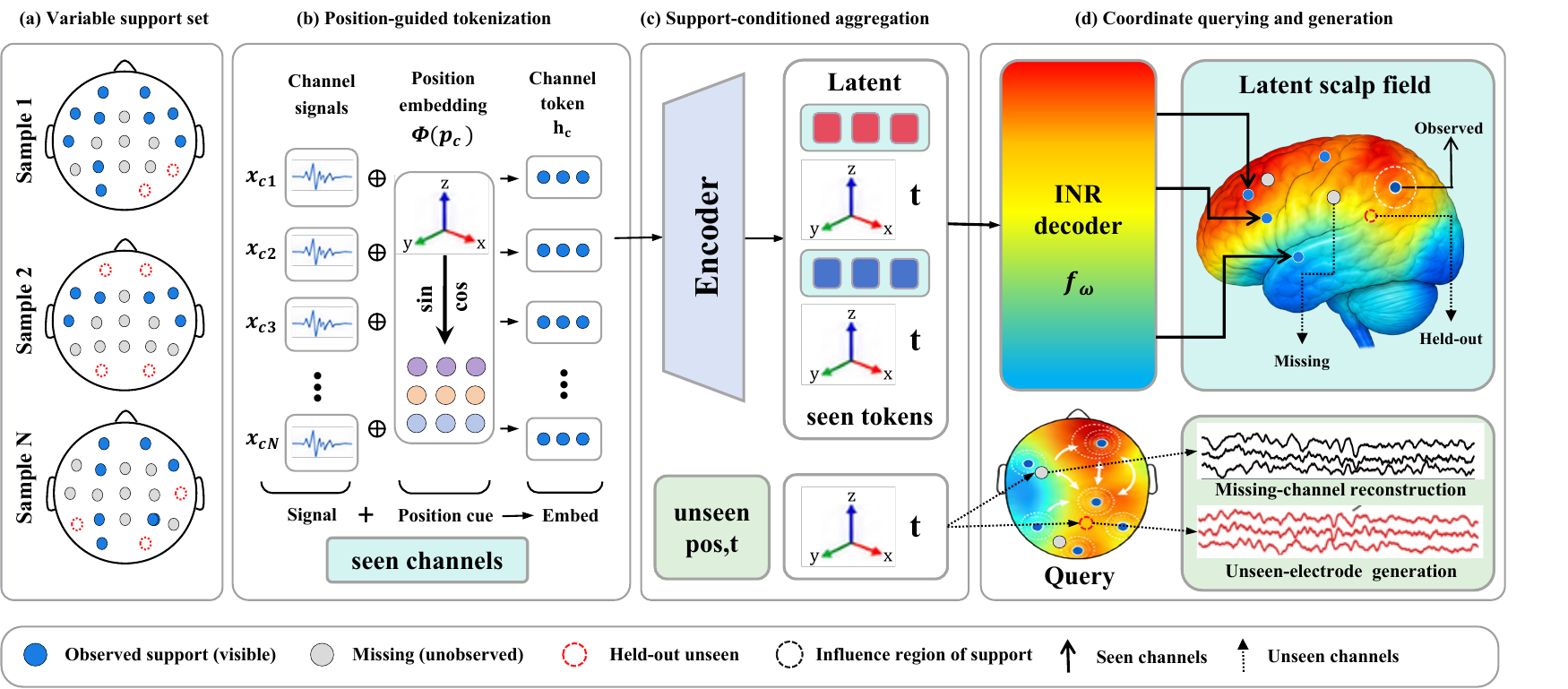}
    \caption{Overview of ScalpINR. Partially observed EEG support channels are encoded into a support-conditioned scalp field, which is queried by a conditional INR decoder to reconstruct missing channels or generate unseen-electrode signals.}

    \label{fig:overview}
\end{figure}

\subsection{Coordinate-queryable conditional INR decoder}

Given the latent representation $z$, the decoder reconstructs EEG signals at arbitrary target electrode locations through a conditional implicit neural representation (INR). Let $p_q \in \mathbb{R}^{3}$ be a query coordinate and $t$ a time index. The decoder predicts the target EEG value by querying a coordinate-conditioned neural field:
\begin{equation}
\hat{x}(p_q,t) = f_{\omega}(z, \phi(p_q), \tau(t)),
\end{equation}
where $\phi(p_q)$ is the embedded query coordinate, $\tau(t)$ is a time embedding, and $f_{\omega}$ is the conditional INR decoder. This formulation decouples visible-channel encoding from target-signal generation: the encoder summarizes what is observed, while the INR decoder synthesizes signals at requested coordinates. As a result, the model is not restricted to restoring a pre-defined output slot, but can generate EEG signals at unseen electrode locations by direct coordinate querying.

In practice, we query the decoder at all target coordinates in the canonical montage to obtain the reconstructed EEG segment
\begin{equation}
\hat{X} = \{\hat{x}(p_q,t)\}_{q=1,t=1}^{C,\;T}.
\end{equation}
When a strict unseen-electrode evaluation protocol is used, a held-out subset of electrode coordinates is never exposed as input during training, and reconstruction at these locations is performed entirely through coordinate querying.

\begin{figure}
    \centering
    \includegraphics[width=\linewidth]{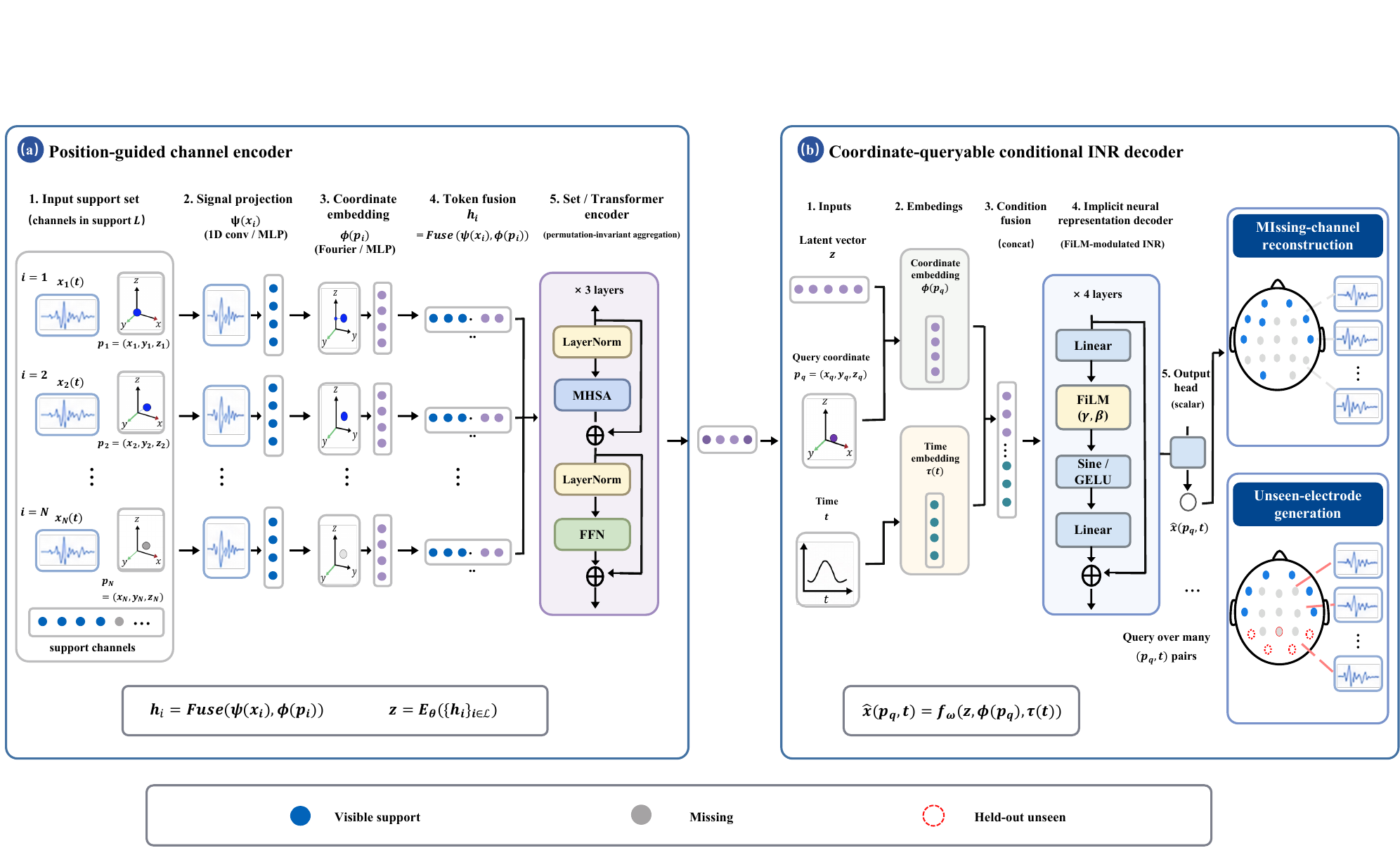}
    \caption{Detailed architecture of ScalpINR. The encoder aggregates position-guided channel tokens into a latent condition, and the coordinate-queryable INR decoder predicts EEG signals at target electrode coordinates.}
    \label{fig:architecture}
\end{figure}

\subsection{Fidelity-preserving channel corruption training}

Recovering missing or unseen channels alone is insufficient for reliable EEGSR. A conditional INR may produce plausible target signals without being sufficiently anchored to the already observed channels, especially under severe missingness. To address this issue, we introduce a fidelity-preserving channel corruption training strategy, whose goal is to strengthen the constraint of the encoded latent representation on the decoder and thereby construct a more stable scalp field consistent with the observed channels.

Specifically, during training we simulate three electrode states:
\begin{itemize}
    \item \textbf{missing electrodes}, whose signals are removed from the input;
    \item \textbf{corrupted electrodes}, whose signals are perturbed to mimic unreliable or degraded observations;
    \item \textbf{visible-but-predicted electrodes}, which remain available in the input but are still included in the prediction target.
\end{itemize}
This mixed-state design goes beyond idealized binary masking and encourages the model to learn a support-consistent reconstruction rule. Missing electrodes teach the decoder to recover unobserved signals, corrupted electrodes improve robustness to unstable channel quality, and visible-but-predicted electrodes prevent the model from treating observed channels as unconstrained pass-through signals. Together, these states force the INR to explain both what is missing and what is already given through a single latent scalp representation.

\subsection{Training objectives}

Our training objective contains two components. First, we apply a reconstruction loss on the target channels:
\begin{equation}
\mathcal{L}_{\mathrm{rec}} = \frac{1}{|\mathcal{Q}|T}\sum_{q \in \mathcal{Q}} \sum_{t=1}^{T} \ell\big(\hat{x}(p_q,t), x(p_q,t)\big),
\end{equation}
where $\mathcal{Q}$ denotes the set of supervised target electrodes and $\ell(\cdot,\cdot)$ is an element-wise regression loss such as $\ell_1$ or $\ell_2$. Second, to preserve consistency with already observed channels, we apply an additional fidelity term on the visible channels selected for prediction:
\begin{equation}
\mathcal{L}_{\mathrm{fid}} = \frac{1}{|\mathcal{L}_{p}|T}\sum_{c \in \mathcal{L}_{p}} \sum_{t=1}^{T} \ell\big(\hat{x}(p_c,t), x_c(t)\big),
\end{equation}
where $\mathcal{L}_{p}\subseteq \mathcal{L}$ denotes the visible-but-predicted channels. The final objective is
\begin{equation}
\mathcal{L} = \mathcal{L}_{\mathrm{rec}} + \lambda \mathcal{L}_{\mathrm{fid}},
\end{equation}
where $\lambda$ balances missing-channel recovery and fidelity preservation.

\subsection{Training and inference protocol}

For random missing-channel reconstruction, we randomly sample visible-channel patterns during training and evaluation, covering both unstructured and structured missingness. For unseen-electrode generation, we define a held-out electrode subset whose coordinates are never used as input during training. At inference time, the trained model first encodes the available channels, then queries the INR decoder at the target electrode coordinates to synthesize the corresponding EEG signals. This unified formulation allows the same framework to handle both random missing-channel reconstruction and strict unseen-electrode generation.

\section{Experiments}

\subsection{Experiment Setup}

\paragraph{Datasets.}
We evaluate ScalpINR on three publicly available EEG datasets, including \textbf{SEED} \citep{zheng2015investigating}, \textbf{AAD}, and \textbf{BCI2000}. These datasets cover emotion-related EEG, auditory-attention EEG, and motor-imagery EEG, providing diverse electrode layouts, signal statistics, and downstream decoding tasks for evaluating both reconstruction fidelity and task utility. Detailed dataset descriptions and preprocessing pipelines are provided in Appendix~\ref{app:datasets}.

\paragraph{Baselines}
We compare our method against representative EEG spatial super-resolution baselines from different modeling families, including 
\textbf{ImputeINR}~\citep{li2025imputeinr}, \textbf{CSDI}~\citep{tashiro2021csdi}, \textbf{ZUNA}~\citep{warner2026zuna}, \textbf{ESTformer}~\citep{li2025estformer}, \textbf{SRGDiff}~\citep{liu2026step}, \textbf{MCMA}~\citep{chen2024multi}. We further include \textbf{spherical spline interpolation} as a classical geometry-based baseline for missing-channel recovery. All methods are trained and evaluated under the same preprocessing pipeline, data splits, and missingness settings. Implementation details are provided in Appendix~\ref{app:baseline}.

\paragraph{Evaluation protocol} test 50 times.
Following prior EEGSR evaluation practice\citep{liu2026step}, we evaluate performance from three complementary perspectives:
\textbf{signal-level} fidelity (NMSE, Pearson correlation, and SNR),
\textbf{spectral/feature-level} fidelity (e.g., STFT power-spectrum error and EEG-FID computed with a frozen encoder),
and \textbf{downstream-level} utility (task accuracy on standard EEG decoding tasks). 

\subsection{Main Results under Random Missing-Channel Reconstruction}

\begin{table*}[t]
\centering
\caption{Random missing-channel reconstruction results on \textbf{SEED} under different support ratios $r$.
Here, $r$ denotes the fraction of visible channels used as model input.
Lower NMSE and higher PCC/SNR indicate better performance.
Best results are in bold and second-best results are underlined.}
\label{tab:seed_recon}
\resizebox{\textwidth}{!}{
\begin{tabular}{lccccccccc}
\toprule
\multirow{2}{*}{Method}
& \multicolumn{3}{c}{$r = 0.5$}
& \multicolumn{3}{c}{$r = 0.25$}
& \multicolumn{3}{c}{$r = 0.125$} \\
\cmidrule(lr){2-4} \cmidrule(lr){5-7} \cmidrule(lr){8-10}
& NMSE $\downarrow$ & PCC $\uparrow$ & SNR $\uparrow$
& NMSE $\downarrow$ & PCC $\uparrow$ & SNR $\uparrow$
& NMSE $\downarrow$ & PCC $\uparrow$ & SNR $\uparrow$ \\
\midrule

Spherical Spline
& 0.70 & 0.64 & 1.96
& 1.14 & 0.53 & -0.14
& 2.06 & 0.40 & -2.71 \\

ZUNA
& 0.97 & 0.10 & 0.23
& 0.98 & 0.07 & 0.21
& 0.99 & 0.00 & -0.01 \\

ImputeINR
& 0.61 & 0.66 & 2.54
& 0.86 & 0.57 & 1.73
& 1.54 & 0.47 & -0.38 \\

CSDI
& 0.75 & 0.59 & 0.75
& 1.26 & 0.48 & -1.06
& 2.38 & 0.35 & -4.16 \\

ESTformer
& 0.53 & 0.68 & 2.95
& 0.60 & 0.63 & 2.32
& 0.69 & 0.57 & 1.69 \\

MCMA
& \underline{0.52} & \underline{0.69} & \underline{3.39}
& \underline{0.57} & \underline{0.65} & \underline{2.59}
& \underline{0.64} & \underline{0.60} & \underline{2.28} \\

SRGDiff
& 0.67 & 0.65 & 2.37
& 1.08 & 0.54 & 0.13
& 1.87 & 0.42 & -2.16 \\

\midrule

ScalpINR
& \textbf{0.48} & \textbf{0.72} & \textbf{3.40}
& \textbf{0.50} & \textbf{0.70} & \textbf{3.14}
& \textbf{0.54} & \textbf{0.67} & \textbf{2.80} \\

\bottomrule
\end{tabular}
}
\end{table*}

\begin{table*}[t]
\centering
\caption{Random missing-channel reconstruction results on \textbf{AAD} under different support ratios $r$.
Here, $r$ denotes the fraction of visible channels used as model input.
Lower NMSE and higher PCC/SNR indicate better performance.
Best results are in bold and second-best results are underlined.}
\label{tab:aad_recon}
\resizebox{\textwidth}{!}{
\begin{tabular}{lcccccccccccc}
\toprule
\multirow{2}{*}{Method}
& \multicolumn{3}{c}{$r = 0.5$}
& \multicolumn{3}{c}{$r = 0.25$}
& \multicolumn{3}{c}{$r = 0.125$}
& \multicolumn{3}{c}{$r = 0.0625$} \\
\cmidrule(lr){2-4} \cmidrule(lr){5-7} \cmidrule(lr){8-10} \cmidrule(lr){11-13}
& NMSE $\downarrow$ & PCC $\uparrow$ & SNR $\uparrow$
& NMSE $\downarrow$ & PCC $\uparrow$ & SNR $\uparrow$
& NMSE $\downarrow$ & PCC $\uparrow$ & SNR $\uparrow$
& NMSE $\downarrow$ & PCC $\uparrow$ & SNR $\uparrow$ \\
\midrule

Spherical Spline
& 0.39 & 0.78 & 4.51
& 0.47 & 0.75 & 3.66
& 0.55 & 0.70 & 2.97
& 0.69 & 0.53 & 1.78 \\

ZUNA
& 0.75 & 0.50 & 1.49
& 0.78 & 0.45 & 1.41
& 0.75 & 0.50 & 1.47
& 0.80 & 0.43 & 1.20 \\

ImputeINR
& 0.37 & \underline{0.80} & \underline{4.96}
& 0.47 & 0.75 & 3.31
& 0.57 & 0.67 & 2.49
& 0.67 & 0.58 & 1.34 \\

CSDI
& 0.41 & 0.75 & 3.19
& 0.57 & 0.62 & 1.92
& 0.68 & 0.57 & 1.06
& 0.77 & 0.46 & 1.42 \\

ESTformer
& \textbf{0.34} & \textbf{0.81} & \textbf{5.04}
& 0.46 & 0.76 & 3.24
& 0.58 & 0.66 & 1.89
& 0.68 & 0.57 & 1.16 \\

MCMA
& 0.38 & \underline{0.80} & 4.60
& \underline{0.40} & \underline{0.78} & \underline{4.25}
& \underline{0.45} & \underline{0.75} & \underline{3.72}
& \underline{0.53} & \underline{0.68} & \underline{2.91} \\

SRGDiff
& 0.37 & 0.79 & 4.92
& 0.49 & 0.74 & 3.43
& 0.59 & 0.68 & 2.61
& 0.72 & 0.50 & 1.58 \\

\midrule

ScalpINR
& \underline{0.36} & \underline{0.80} & 4.92
& \textbf{0.37} & \textbf{0.79} & \textbf{4.68}
& \textbf{0.39} & \textbf{0.78} & \textbf{4.48}
& \textbf{0.41} & \textbf{0.76} & \textbf{4.19} \\

\bottomrule
\end{tabular}
}
\end{table*}

\begin{table*}[t]
\centering
\caption{Random missing-channel reconstruction results on \textbf{BCI2000} under different support ratios $r$.
Here, $r$ denotes the fraction of visible channels used as model input.
Lower NMSE and higher PCC/SNR indicate better performance.
Best results are in bold and second-best results are underlined.}
\label{tab:bci2000_recon}
\resizebox{\textwidth}{!}{
\begin{tabular}{lccccccccc}
\toprule
\multirow{2}{*}{Method}
& \multicolumn{3}{c}{$r = 0.5$}
& \multicolumn{3}{c}{$r = 0.25$}
& \multicolumn{3}{c}{$r = 0.125$} \\
\cmidrule(lr){2-4} \cmidrule(lr){5-7} \cmidrule(lr){8-10}
& NMSE $\downarrow$ & PCC $\uparrow$ & SNR $\uparrow$
& NMSE $\downarrow$ & PCC $\uparrow$ & SNR $\uparrow$
& NMSE $\downarrow$ & PCC $\uparrow$ & SNR $\uparrow$ \\
\midrule

Spherical Spline
& 0.27 & 0.86 & 6.27
& 0.41 & 0.80 & 4.29
& 0.68 & 0.71 & 2.09 \\

ZUNA
& 0.96 & 0.20 & 0.33
& 0.99 & 0.03 & 0.09
& 0.99 & 0.01 & 0.01 \\

ImputeINR
& 0.23 & 0.88 & 6.68
& 0.35 & 0.85 & 5.62
& 0.68 & 0.71 & 1.64 \\

CSDI
& 0.33 & 0.82 & 5.16
& 0.51 & 0.77 & 3.68
& 0.82 & 0.67 & 1.36 \\

ESTformer
& 0.24 & 0.87 & 6.39
& 0.48 & 0.81 & 3.20
& 0.70 & 0.73 & 1.56 \\

MCMA
& \underline{0.21} & \textbf{0.89} & \underline{7.07}
& \underline{0.25} & \underline{0.87} & \underline{6.33}
& \underline{0.31} & \underline{0.83} & \underline{5.32} \\

SRGDiff
& 0.25 & 0.87 & 6.46
& 0.37 & 0.83 & 5.01
& 0.71 & 0.69 & 2.14 \\

\midrule

ScalpINR
& \textbf{0.20} & \textbf{0.89} & \textbf{7.19}
& \textbf{0.23} & \textbf{0.88} & \textbf{6.68}
& \textbf{0.25} & \textbf{0.86} & \textbf{6.13} \\

\bottomrule
\end{tabular}
}
\end{table*}

Under the random missing-channel reconstruction setting, we sample visible channels with both missing rates and missing positions vary across samples. Detailed missingness descriptions are provided in Appendix~\ref{app:missingness}.

\paragraph{Signal-level reconstruction performance.}
We first evaluate random missing-channel reconstruction using waveform-level metrics, including NMSE, PCC, and SNR. As shown in Tables~\ref{tab:seed_recon}--\ref{tab:bci2000_recon}, ScalpINR achieves the strongest overall performance across the three EEG datasets, with particularly clear gains under low support ratios. On SEED, when only $r=0.125$ channels are visible, ScalpINR reduces NMSE from $0.64$ to $0.54$ and improves PCC/SNR from $0.60$/$2.28$ to $0.67$/$2.80$ over the strongest baseline. On BCI2000, ScalpINR further reduces NMSE from $0.31$ to $0.25$ and improves SNR from $5.32$ dB to $6.13$ dB at $r=0.125$. On AAD, although ESTformer performs slightly better under the mild $r=0.5$ setting, ScalpINR becomes consistently superior as the support ratio decreases, achieving $0.41$ NMSE, $0.76$ PCC, and $4.19$ dB SNR at the extremely sparse $r=0.0625$ setting. These results indicate that support-conditioned coordinate-queryable reconstruction is especially effective when the visible EEG channels are sparse and randomly distributed.

\paragraph{Spectral reconstruction analysis.}
We further evaluate frequency-domain reconstruction to examine whether the recovered signals preserve EEG spectral structure beyond point-wise waveform similarity. As reported in Table~\ref{tab:spectral}, ScalpINR consistently achieves the lowest spectral error across datasets and support ratios. On SEED, ScalpINR reduces the spectral error from $19$ to $15$ at $r=0.25$ and from $27$ to $21$ at $r=0.125$ compared with the strongest baseline. On BCI2000, the advantage becomes clearer under severe sparsity, where ScalpINR reduces the error from $19$ to $13$ at $r=0.125$. On AAD, ScalpINR obtains the largest improvement under the most challenging setting, reducing the spectral error from $22$ to $13$ at $r=0.0625$. These results suggest that ScalpINR not only improves signal-level reconstruction accuracy, but also better preserves the frequency-domain structure of EEG signals under sparse and variable channel observations.

\begin{table*}[t]
\centering
\caption{Frequency-domain reconstruction results under different support ratios $r$.
Here, $r$ denotes the fraction of visible channels used as model input.
Lower values indicate better spectral preservation.
Best results are in bold and second-best results are underlined.}
\label{tab:spectral}
\resizebox{\textwidth}{!}{
\begin{tabular}{lccccccccc}
\toprule
\multirow{2}{*}{Method}
& \multicolumn{2}{c}{SEED}
& \multicolumn{3}{c}{BCI2000}
& \multicolumn{4}{c}{AAD} \\
\cmidrule(lr){2-3} \cmidrule(lr){4-6} \cmidrule(lr){7-10}
& $r=0.25$ & $r=0.125$
& $r=0.5$ & $r=0.25$ & $r=0.125$
& $r=0.5$ & $r=0.25$ & $r=0.125$ & $r=0.0625$ \\
\midrule

Spherical Spline
& 204 & 336
& 74 & 91 & 138
& 45 & 91 & 186 & 284 \\

ImputeINR
& 127 & 202
& 41 & 71 & 105
& 63 & 89 & 159 & 238 \\

CSDI
& 76 & 152
& 67 & 83 & 124
& 35 & 54 & 109 & 143 \\

ZUNA
& 238 & 329
& 95 & 116 & 171
& 62 & 120 & 243 & 331 \\

ESTformer
& 19 & 27
& 9 & 21 & 37
& 8 & 9 & 28 & 49 \\

SRGDiff
& 65 & 65
& 13 & 16 & 19
& 6 & 11 & 18 & 22 \\

CAFE
& 137 & 216
& 80 & 94 & 152
& 57 & 98 & 206 & 314 \\

\midrule

Ours
& \textbf{15} & \textbf{21}
& \textbf{9} & \textbf{11} & \textbf{13}
& \textbf{5} & \textbf{9} & \textbf{11} & \textbf{13} \\

\bottomrule
\end{tabular}
}
\end{table*}

\paragraph{Downstream decoding performance.}
To verify whether the reconstructed signals preserve task-discriminative information, we evaluate downstream decoding performance using reconstructed EEG signals under different support ratios. The downstream classifier is trained on ground-truth HD signals from the training split and evaluated on reconstructed HD signals from the test split. As shown in Tables~\ref{tab:downstream_seed}--\ref{tab:downstream_aad}, ScalpINR achieves the best or tied-best downstream performance across all evaluated datasets and support ratios. On SEED, ScalpINR improves the accuracy from $0.37$ to $0.40$ at $r=0.25$ and from $0.35$ to $0.36$ at $r=0.125$ over the strongest baseline, reaching the corresponding GT-reference performance under severe sparsity. On BCI2000, ScalpINR achieves $0.69$ and $0.64$ accuracy at $r=0.25$ and $r=0.125$, outperforming the strongest baseline by $0.01$ and $0.02$, respectively. On AAD, ScalpINR improves the accuracy from $0.53$ to $0.57$ at $r=0.5$ and maintains the best performance under lower support ratios. These results indicate that the proposed coordinate-queryable reconstruction not only improves waveform and spectral fidelity, but also better preserves task-relevant neural information for downstream decoding.

\begin{table*}[t]
\centering
\caption{Downstream decoding accuracy on \textbf{SEED} using reconstructed EEG signals under different support ratios $r$.
The downstream classifier is trained on ground-truth HD signals and evaluated on reconstructed HD test signals.
The row ``GT-HD'' reports the reference accuracy obtained using real HD signals under the corresponding evaluation setting.
Higher values indicate better task utility.
Tied best values are bolded, and second-best values are underlined.}
\label{tab:downstream_seed}
\resizebox{0.7\textwidth}{!}{
\begin{tabular}{lcccc}
\toprule
Method & Full HD & $r=0.5$ & $r=0.25$ & $r=0.125$ \\
\midrule
GT-HD            & 0.45 & 0.42 & 0.40 & 0.36 \\
\midrule
Spherical Spline & --   & 0.33 & 0.33 & 0.33 \\
ZUNA             & --   & 0.33 & 0.33 & 0.33 \\
ImputeINR        & --   & \underline{0.40} & 0.34 & 0.34 \\
CSDI             & --   & 0.35 & 0.33 & 0.33 \\
ESTformer        & --   & 0.39 & \underline{0.37} & \underline{0.35} \\
MCMA             & --   & 0.35 & 0.34 & 0.34 \\
SRGDiff          & --   & 0.36 & 0.33 & 0.33 \\
\midrule
ScalpINR         & --   & \textbf{0.41} & \textbf{0.40} & \textbf{0.36} \\
\bottomrule
\end{tabular}
}
\end{table*}

\begin{table*}[t]
\centering
\caption{Downstream decoding accuracy on \textbf{BCI2000} using reconstructed EEG signals under different support ratios $r$.
The downstream classifier is trained on ground-truth HD signals and evaluated on reconstructed HD test signals.
The row ``GT-HD'' reports the reference accuracy obtained using real HD signals under the corresponding evaluation setting.
Higher values indicate better task utility.
Tied best values are bolded, and second-best values are underlined.}
\resizebox{0.7\textwidth}{!}{
\begin{tabular}{lcccc}
\toprule
Method & Full HD & $r=0.5$ & $r=0.25$ & $r=0.125$ \\
\midrule
GT-HD            & 0.75 & 0.72 & 0.70 & 0.66 \\
\midrule
Spherical Spline & --   & 0.61 & 0.53 & 0.51 \\
ZUNA             & --   & 0.50 & 0.50 & 0.50 \\
ImputeINR        & --   & 0.69 & 0.66 & 0.59 \\
CSDI             & --   & 0.53 & 0.50 & 0.50 \\
ESTformer        & --   & \textbf{0.71} & 0.67 & 0.60 \\
MCMA             & --   & 0.68 & \underline{0.68} & \underline{0.62} \\
SRGDiff          & --   & 0.56 & 0.50 & 0.50 \\
\midrule
ScalpINR         & --   & \textbf{0.71} & \textbf{0.69} & \textbf{0.64} \\
\bottomrule
\end{tabular}
}
\end{table*}

\begin{table*}[t]
\centering
\caption{Downstream decoding accuracy on \textbf{AAD} using reconstructed EEG signals under different support ratios $r$.
The downstream classifier is trained on ground-truth HD signals and evaluated on reconstructed HD test signals.
The row ``GT-HD'' reports the reference accuracy obtained using real HD signals under the corresponding evaluation setting.
Higher values indicate better task utility.
Tied best values are bolded, and second-best values are underlined.}
\label{tab:downstream_aad}
\resizebox{0.7\textwidth}{!}{
\begin{tabular}{lcccc}
\toprule
Method & Full HD & $r=0.5$ & $r=0.25$ & $r=0.125$ \\
\midrule
GT-HD            & 0.63 & 0.59 & 0.53 & 0.51 \\
\midrule
Spherical Spline & --   & 0.50 & 0.50 & 0.50 \\
ZUNA             & --   & 0.50 & 0.50 & 0.50 \\
ImputeINR        & --   & 0.52 & 0.50 & 0.50 \\
CSDI             & --   & 0.50 & 0.50 & 0.50 \\
ESTformer        & --   & 0.51 & \underline{0.51} & 0.50 \\
MCMA             & --   & 0.52 & \underline{0.51} & \underline{0.51} \\
SRGDiff          & --   & \underline{0.53} & 0.50 & 0.50 \\
\midrule
ScalpINR         & --   & \textbf{0.57} & \textbf{0.53} & \textbf{0.52} \\
\bottomrule
\end{tabular}
}
\end{table*}

\begin{table*}[t]
\centering
\caption{Strict unseen-electrode generation results under different support ratios $r$, evaluated only on permanently held-out electrodes.
Here, $r$ denotes the fraction of non-held-out electrodes used as visible support, and $r=1$ means that all non-held-out electrodes are available while the held-out electrodes are used only as query targets.
We report mean values across multiple random hold-out splits.
Lower NMSE and higher PCC/SNR indicate better performance.
Best results are in bold and second-best results are underlined.}
\label{tab:unseen_gen_sr}
\resizebox{\textwidth}{!}{
\begin{tabular}{llccccccccc}
\toprule
\multirow{2}{*}{$r$} & \multirow{2}{*}{Method} 
& \multicolumn{3}{c}{SEED (held-out only)} 
& \multicolumn{3}{c}{AAD (held-out only)} 
& \multicolumn{3}{c}{BCI2000 (held-out only)} \\
\cmidrule(lr){3-5} \cmidrule(lr){6-8} \cmidrule(lr){9-11}
& 
& NMSE $\downarrow$ & PCC $\uparrow$ & SNR $\uparrow$
& NMSE $\downarrow$ & PCC $\uparrow$ & SNR $\uparrow$
& NMSE $\downarrow$ & PCC $\uparrow$ & SNR $\uparrow$ \\
\midrule

\multirow{3}{*}{1}
& Spherical Spline
& \underline{0.29} & \underline{0.84} & \underline{5.78}
& \underline{0.16} & \underline{0.94} & \underline{8.22}
& \underline{0.18} & \underline{0.90} & \underline{7.52} \\
& ZUNA
& 0.85 & 0.32 & 0.77
& 0.70 & 0.51 & 1.66
& 0.90 & 0.24 & 0.38 \\
& ScalpINR
& \textbf{0.26} & \textbf{0.86} & \textbf{6.09}
& \textbf{0.10} & \textbf{0.95} & \textbf{10.34}
& \textbf{0.16} & \textbf{0.95} & \textbf{7.93} \\
\midrule

\multirow{3}{*}{0.5}
& Spherical Spline
& \underline{0.73} & \underline{0.63} & \underline{1.87}
& \underline{0.42} & \underline{0.80} & \underline{4.63}
& \underline{0.28} & \underline{0.87} & \underline{6.31} \\
& ZUNA
& 0.93 & 0.12 & 0.31
& 0.77 & 0.47 & 1.24
& 0.95 & 0.21 & 0.39 \\
& ScalpINR
& \textbf{0.34} & \textbf{0.83} & \textbf{4.78}
& \textbf{0.22} & \textbf{0.94} & \textbf{6.58}
& \textbf{0.16} & \textbf{0.94} & \textbf{7.94} \\
\midrule

\multirow{3}{*}{0.25}
& Spherical Spline
& 1.30 & \underline{0.51} & -0.72
& \underline{0.68} & \underline{0.75} & \underline{3.58}
& \textbf{0.43} & \underline{0.78} & \textbf{4.06} \\
& ZUNA
& \underline{1.05} & 0.06 & \underline{0.17}
& 0.81 & 0.45 & 1.36
& 0.92 & 0.04 & 0.11 \\
& ScalpINR
& \textbf{0.58} & \textbf{0.75} & \textbf{2.32}
& \textbf{0.35} & \textbf{0.90} & \textbf{4.54}
& \underline{0.49} & \textbf{0.86} & \underline{3.02} \\
\midrule

\multirow{3}{*}{0.125}
& Spherical Spline
& 2.41 & \underline{0.37} & -3.16
& 0.87 & \underline{0.69} & \underline{2.84}
& \underline{0.72} & \underline{0.69} & \underline{1.87} \\
& ZUNA
& \underline{1.21} & 0.01 & \underline{-1.22}
& \underline{0.79} & 0.43 & 1.18
& 0.95 & 0.01 & 0.08 \\
& ScalpINR
& \textbf{0.70} & \textbf{0.69} & \textbf{1.50}
& \textbf{0.47} & \textbf{0.83} & \textbf{3.71}
& \textbf{0.62} & \textbf{0.86} & \textbf{2.02} \\
\bottomrule
\end{tabular}
}
\end{table*}
\subsection{Low-Density Support: Joint Reconstruction and Extrapolation}

We further evaluate ScalpINR under a stricter low-density support setting that jointly considers sparse-support reconstruction and unseen-electrode generation. 
For each dataset, approximately $10\%$ of electrodes are permanently designated as \emph{unseen electrodes}. 
These electrodes are never used as input support during training and are only queried as target locations during evaluation. 
The support ratio $r$ is then defined over the remaining non-held-out electrodes. 
In particular, $r=1$ means that all non-held-out electrodes are available as support, while the held-out electrodes are still used only as query targets. 
Detailed missingness descriptions are provided in Appendix~\ref{app:missingness}. 
This protocol directly examines whether a model can infer a coordinate-queryable scalp field from sparse observed support and extrapolate to electrode locations that are not exposed as support during training.

As shown in Table~\ref{tab:unseen_gen_sr}, ScalpINR achieves the best performance in most strict held-out settings, especially on SEED and AAD. 
Under the clean held-out setting with $r=1$, ScalpINR reduces the NMSE on AAD from $0.16$ to $0.10$ compared with the strongest baseline, $ 37.5\%$ relative reduction. 
At the same time, the SNR improves from $8.22$ dB to $10.34$ dB, yielding $2.12 \ \mathrm{dB}$ absolute improvement. 
This result directly supports the coordinate-queryable design of ScalpINR, since the evaluated target electrodes are never observed as support during training.

The advantage remains clear when the support becomes sparse. 
For example, on AAD with $r=0.5$, ScalpINR reduces NMSE from $0.42$ to $0.22$, giving $ 47.6\% $ relative reduction over spherical spline interpolation. 
On SEED with $r=0.5$, ScalpINR also improves PCC from $0.63$ to $0.83$ and SNR from $1.87$ dB to $4.78$ dB, showing that the learned conditional scalp field can generalize beyond interpolation when the visible support is limited. 
On BCI2000, ScalpINR performs best under $r=1$, $r=0.5$, and $r=0.125$, while spherical spline remains competitive at $r=0.25$ in NMSE and SNR. 
This suggests that very low-channel or strongly structured layouts may provide less spatial redundancy for neural field extrapolation, but ScalpINR still maintains stronger correlation with the held-out signals. 

\subsection{Qualitative Visualization}

\begin{figure*}[t]
\centering
\includegraphics[width=\textwidth]{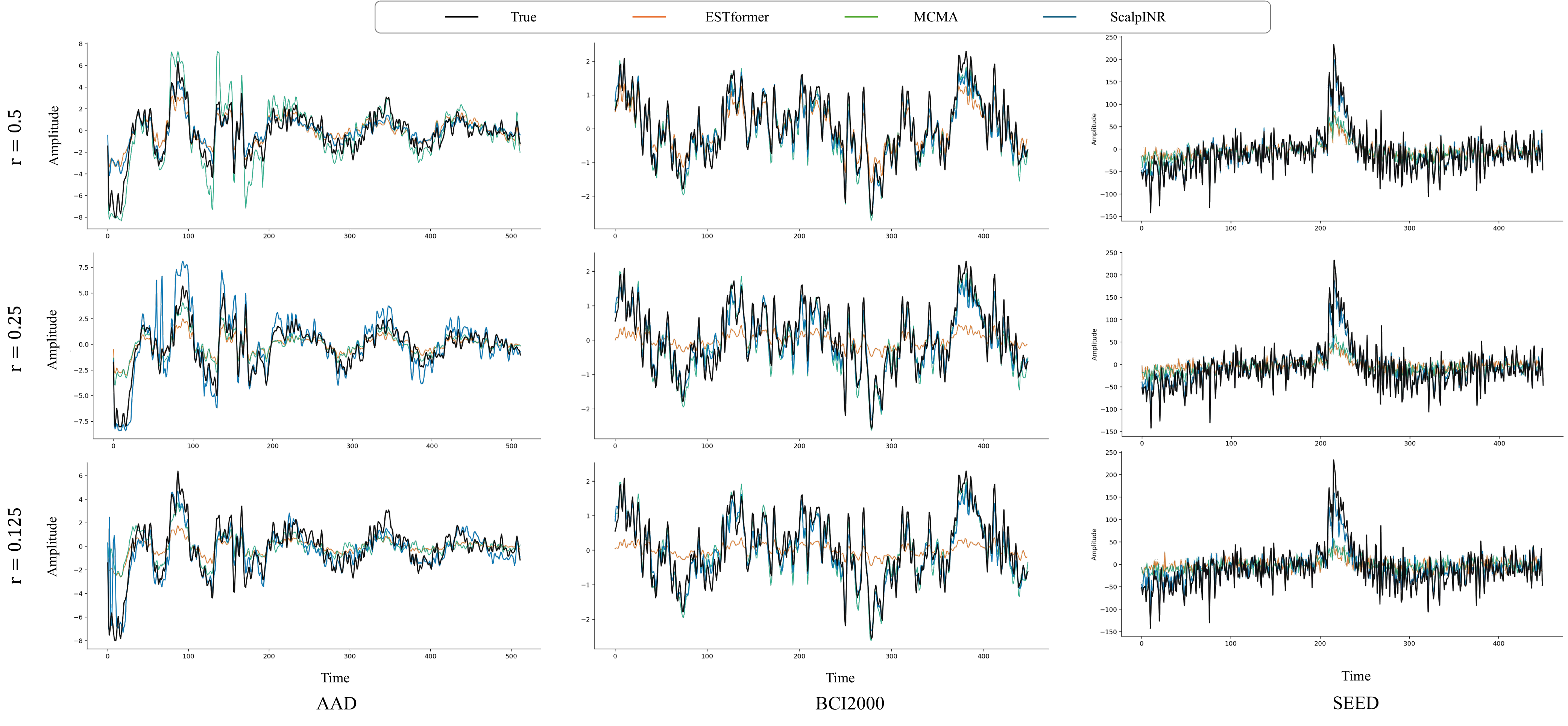}
\caption{Qualitative waveform comparison under random missing-channel reconstruction. 
Rows correspond to different support ratios $r$, and curves show the ground-truth signal and reconstructions produced by ESTformer, MCMA, and ScalpINR. 
ScalpINR better follows the temporal morphology of the ground truth, especially under lower support ratios where fixed-layout baselines exhibit stronger amplitude attenuation or local waveform distortion.}
\label{fig:vis_random}
\end{figure*}

We provide qualitative waveform comparisons for both random missing-channel reconstruction and strict unseen-electrode generation. 
Figure~\ref{fig:vis_random} shows representative reconstruction examples under random missing-channel settings with different support ratios. 
Compared with ESTformer and MCMA, ScalpINR more closely follows the ground-truth temporal morphology across both moderate and severe sparsity. 
This advantage is especially visible when the support ratio decreases to $r=0.25$ and $r=0.125$, where baseline methods tend to produce attenuated, shifted, or locally distorted waveforms, while ScalpINR better preserves the amplitude variation and temporal peaks of the reference signal.

\begin{figure*}[t]
\centering
\includegraphics[width=\textwidth]{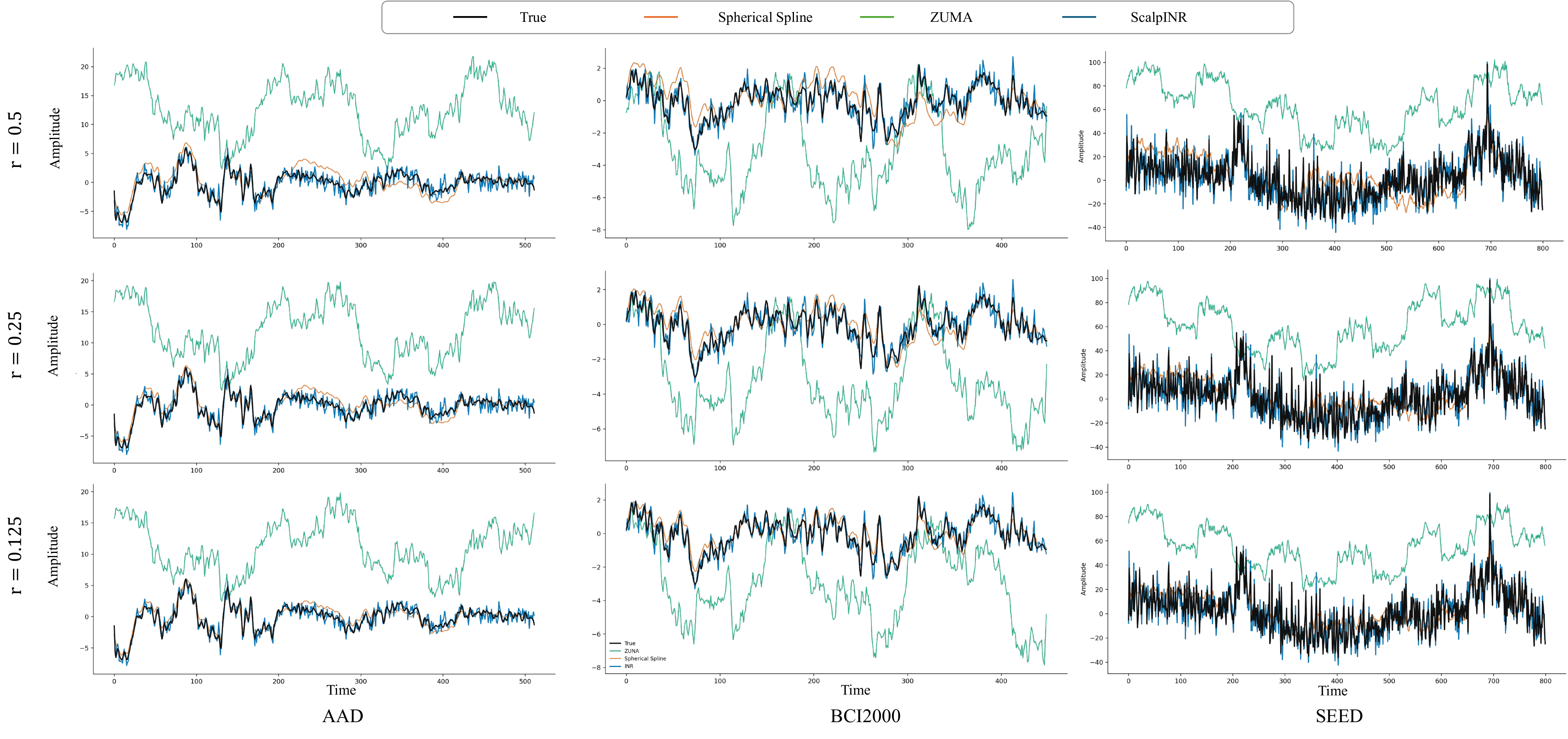}
\caption{Qualitative waveform comparison under strict unseen-electrode generation. 
The target electrodes are permanently held out from the support set during training and are queried only during evaluation. 
Rows correspond to different support ratios $r$, and columns show representative examples from AAD, BCI2000, and SEED. 
Compared with spherical spline interpolation and ZUNA, ScalpINR produces waveforms that are more consistent with the ground truth, demonstrating its ability to synthesize EEG signals at unseen electrode coordinates.}
\label{fig:vis_unseen}
\end{figure*}

Figure~\ref{fig:vis_unseen} further visualizes strict unseen-electrode generation, where the queried target electrodes are never used as support during training. 
Under this more challenging setting, spherical spline interpolation can capture coarse trends but often over-smooths local waveform details, while ZUNA shows clear deviations from the ground truth in several cases. 
In contrast, ScalpINR remains closely aligned with the real signals across SEED, AAD, and BCI2000, even under low support ratios. 
These qualitative results support the quantitative findings that ScalpINR learns a support-conditioned coordinate-queryable scalp field rather than merely restoring a fixed output layout.

\subsection{Efficiency Analysis}

We further compare the computational cost and inference latency of ScalpINR with representative baselines. 
As shown in Table~\ref{tab:comp_efficiency}, ScalpINR contains only $2.563$M trainable parameters and requires $3.935$ GFLOPs under the SEED input shape of $N \times 62 \times 800$ with support ratio $r=0.125$. 
Although spherical spline interpolation is non-parametric and slightly faster, its reconstruction accuracy is substantially lower than ScalpINR, especially under sparse support and strict unseen-electrode settings. 
Among learnable reconstruction models, ScalpINR achieves the lowest inference latency, requiring only $2.591$ ms per sample. 
Compared with diffusion-based methods, ScalpINR is much more efficient, reducing runtime from $92.541$ ms for SRGDiff to $2.591$ ms, corresponding to a $97.2\%$ latency reduction. 
It is also faster than ESTformer, MCMA, ImputeINR, and CAFE, while maintaining competitive model size and moderate GFLOPs.

\begin{table*}[t]
\centering
\caption{Computational cost and inference latency under the SEED input shape $N \times 62 \times 800$ with support ratio $r=0.125$. 
For coordinate-queryable methods, runtime includes both support encoding and querying all target electrodes.
Runtime is measured as the average per-sample inference latency on an RTX 4090 over 100 runs.
Spherical Spline is a non-parametric interpolation method, and thus has no trainable parameters or GFLOPs.}
\label{tab:comp_efficiency}
\small
\begin{tabular}{lccc}
\toprule
\textbf{Method} 
& \textbf{\#Params (M)} 
& \textbf{GFLOPs} 
& \textbf{Runtime (ms)} \\
\midrule
Spherical Spline
& N/A & N/A & 1.842 \\

ZUNA
& 380 & 46.218 & 118.734 \\

ImputeINR
& 1.247 & 2.684 & 7.913 \\

CSDI
& 4.414 & 38.572 & 286.441 \\

ESTformer
& 12.111 & 4.302 & 5.039 \\

MCMA
& 4.836 & 4.917 & 6.284 \\

SRGDiff
& 20.421 & 5.439 & 92.541 \\

\midrule
ScalpINR
& 2.563 & 3.935 &  2.591\\
\bottomrule
\end{tabular}
\end{table*}






\subsection{Ablation Studies}

We conduct ablations on the main components of the proposed framework, including the position-guided encoder, the coordinate-queryable conditional INR decoder, and the fidelity-preserving channel corruption training strategy. The results verify that each component contributes to the final performance and that the full model achieves the best balance between reconstruction fidelity and generalization to sparse and unseen targets.
\begin{table*}[t]
\centering
\caption{Ablation studies on SEED under different support ratios.
Lower NMSE and higher PCC/SNR indicate better reconstruction performance.}
\label{tab:ablation_seed}
\resizebox{\textwidth}{!}{
\begin{tabular}{lcccccccccccc}
\toprule
\multirow{2}{*}{Variant}
& \multicolumn{3}{c}{SR = 1}
& \multicolumn{3}{c}{SR = 0.5}
& \multicolumn{3}{c}{SR = 0.25}
& \multicolumn{3}{c}{SR = 0.125} \\
\cmidrule(lr){2-4}
\cmidrule(lr){5-7}
\cmidrule(lr){8-10}
\cmidrule(lr){11-13}
& NMSE $\downarrow$ & PCC $\uparrow$ & SNR $\uparrow$
& NMSE $\downarrow$ & PCC $\uparrow$ & SNR $\uparrow$
& NMSE $\downarrow$ & PCC $\uparrow$ & SNR $\uparrow$
& NMSE $\downarrow$ & PCC $\uparrow$ & SNR $\uparrow$ \\
\midrule
w/o mask
& 0.27 & 0.86 & 5.73
& 0.39 & 0.81 & 4.20
& 0.67 & 0.69 & 1.75
& 0.88 & 0.63 & 0.55 \\

w/o coordinate-queryable INR decoder
& 0.27 & 0.85 & 5.69
& 0.38 & 0.81 & 4.11
& 0.64 & 0.71 & 1.91
& 0.79 & 0.66 & 0.96 \\

w/o position-guided encoder
& 0.28 & 0.85 & 5.55
& 0.37 & 0.82 & 4.33
& 0.61 & 0.73 & 2.08
& 0.75 & 0.67 & 1.23 \\
\midrule
full model
& \textbf{0.26} & \textbf{0.86} & \textbf{6.09}
& \textbf{0.34} & \textbf{0.83} & \textbf{4.78}
& \textbf{0.58} & \textbf{0.75} & \textbf{2.32}
& \textbf{0.70} & \textbf{0.69} & \textbf{1.50} \\
\bottomrule
\end{tabular}
}
\end{table*}

\section{Conclusion}

In this paper, we presented ScalpINR, a support-conditioned implicit neural representation framework for EEG spatial super-resolution under random channel missingness. Instead of learning a fixed low-to-high channel mapping tied to a predefined electrode layout, ScalpINR reformulates EEGSR as coordinate-queryable scalp field reconstruction from partially observed support channels. By combining a position-guided channel encoder with a conditional INR decoder, the proposed framework can reconstruct missing channels within the canonical montage and synthesize signals at previously unseen electrode coordinates. We further introduced a fidelity-preserving channel corruption training strategy to strengthen consistency between the inferred scalp field and the observed support channels. Extensive experiments on SEED, AAD, and BCI2000 show that ScalpINR consistently improves signal-level reconstruction, better preserves frequency-domain structure, and yields more task-useful reconstructed signals. In particular, the strict held-out-electrode evaluation demonstrates that ScalpINR can generate EEG signals at electrode locations never exposed as support during training, highlighting its potential for flexible and robust EEG acquisition under variable channel availability. Future work will extend this coordinate-queryable reconstruction paradigm to cross-device montage transfer and more heterogeneous real-world EEG recording conditions.

\small
\bibliographystyle{plainnat}
\bibliography{ai3d_arxiv}

\newpage
\appendix













\end{document}